\documentclass{revtex4}
\usepackage{amssymb}
\usepackage{amsfonts}
\usepackage{amsmath}

\input{tcilatex}

\begin{document}

\preprint{}
\title{Revisiting the charged BTZ metric in nonlinear electrodynamics}
\author{S. Habib Mazharimousavi}
\email{habib.mazhari@emu.edu.tr}
\author{M. Halilsoy}
\email{mustafa.halilsoy@emu.edu.tr}
\author{T. Tahamtan}
\email{tayabeh.tahamtan@emu.edu.tr}
\affiliation{Department of Physics, Eastern Mediterranean University, G. Magusa, north
Cyprus, Mersin 10 - Turkey, Tel.: +90 392 6301067; fax: +90 3692 365 1604.}
\keywords{Black holes, Nonlinear electrodynamics, Exact solution. }

\begin{abstract}
\textbf{Abstract:} In contrast to its chargeless version the charged
Banados, Taitelboim and Zanelli (BTZ) metric in linear Maxwell
electromagnetism is known to be singular at $r=0$. We show, by employing
nonlinear electrodynamics that one obtains charged, extension of the BTZ
metric with regular electric field. This we do by choosing a logarithmic
Lagrangian for the nonlinear electrodynamics. A Theorem is proved on the
existence of electric black holes and combining this results with a duality
principle disproves the existence of magnetic black holes in $2+1-$%
dimensions.
\end{abstract}

\pacs{PACS number}
\maketitle

\section{Introduction}

Since the seminal work of Born and Infeld (BI) \cite{1} nonlinear
electrodynamics (NED) has attracted much interest in different branches of
physics ranging from field theory to string theory. The main idea in
replacing the linear Maxwell's theory with the NED, is to eliminate
divergences due to Coulomb charges. This strategy worked well to a great
extend but the intricate structure of the latter created different problems.
The simplest application of the problem may be considered in connection with
the Hydrogen atom which is drastically modified in the presence of the BI
electrodynamics \cite{2}. Not only the BI version but different types of NED
also found considerable rooms of application in general relativity. The
expectation in doing all this is similar: Can we eliminate divergences that
arise in classical gravitation due to the attractive power of gravity? In
the case of black holes this amounts to finding regular black hole solutions
in general relativity, not only in higher but in lower dimensions as well.

Banados, Teitelboin and Zanelli (BTZ) gave an example of black hole solution
in $2+1-$dimensions whose source consists of mass ($M$) and a cosmological
constant ($\Lambda $) \cite{3}. Although this solution was free of
singularity, the inclusion of electric charge ($Q$) in the problem turned it
singular at $r=0$ \cite{4}. The question that may be raised naturally is:
Can we construct a regular, charged BTZ spacetime by incorporating a NED
instead of the linear electromagnetism? This is precisely the task that we
undertake in this paper and our results can be summarized as follows: The
nonlinear electric field is perfectly regular but the spacetime preserves
its singularity at $r=0$. Before assuming a particular NED we prove a
Theorem that may be significant in connection with NED and occurrence of
black holes: In any circularly symmetric metric ansatz in $2+1-$dimensions
with pure electric field in NED a black hole solution necessarily forms. As
an example the NED we make the choice of a logarithmic Lagrangian \cite{5}
given by $\mathcal{L}\left( F\right) =-\frac{1}{\beta ^{2}}\ln \left(
1+\beta ^{2}F\right) $, in which $F=F_{\mu \nu }F^{\ \mu \nu }$ is the
Maxwell invariant with the constant NED parameter $\beta $. In the limit $%
\beta \rightarrow 0\mathcal{\ }$we recover the linear Maxwell Lagrangian $%
\mathcal{L}\left( F\right) =-F$, whereas for $\beta \rightarrow \infty $ NED
vanishes and the problem reduces to that of uncharged BTZ. Prior to this we
explain first the implication of such a Lagrangian in a flat spacetime. This
derives the electric field and its potential in $2+1-$dimensions in a closed
form. To the first order in charge ($Q$) the potential $V\sim Q\ln r$ ,
coincides with that of charged BTZ, as it should. The contribution from the
higher order terms in $Q$ remarkably eliminates the singularity, and yields
a regular electric field given by $E=\frac{1}{\sqrt{2}\beta }$ as $%
r\rightarrow 0$. Next, we consider the mass and cosmological constant
together with the logarithmic Lagrangian of NED in a curved space. Let us
remark that although we give the logarithmic Lagrangian as an example our
Theorem is applicable for any NED theories. By applying a coordinate
transformation in the $\left( t,\varphi \right) $ plane we introduce
rotation to the metric and generate magnetic field $\overrightarrow{B}$ from
the electric field in accordance with the Maxwell equations. A duality
principle introduced before \cite{6} employed to obtain a magnetic solution
from an electric solution and vice versa.

By using the aforementioned theorem together with the duality principle
defined a la' \cite{6} we show at the same time the absence of black hole
solution in the pure magnetic solutions.

\section{A theorem in $2+1-$dimensions}

We consider the $2+1-$dimensional action ($G=c=1)$ as%
\begin{equation}
S=\frac{1}{16\pi }\int dx^{3}\sqrt{-g}\left( R-\frac{1}{3}\Lambda +\mathcal{L%
}\left( F\right) \right) ,
\end{equation}%
in which $\mathcal{L}\left( F\right) $ is an arbitrary function of Maxwell
invariant $F=F_{\mu \nu }F^{\mu \nu }$ with $F_{\mu \nu }=\partial _{\mu
}A_{\nu }-\partial _{\nu }A_{\mu }$ and the cosmological constant $\Lambda $%
. Here we start with a static, circularly symmetric line element given by 
\begin{equation}
ds^{2}=-f\left( r\right) dt^{2}+\frac{dr^{2}}{f\left( r\right) }%
+r^{2}d\varphi ^{2},
\end{equation}%
Variation of the action with respect to $A_{\mu }$ yields the nonlinear
Maxwell equation%
\begin{equation}
d\left( ^{\star }\mathbf{F}\mathcal{L}_{F}\right) =0
\end{equation}%
where $^{\star }\mathbf{F}$ is the dual of the Maxwell $2-$form $\mathbf{F}$
and $\mathcal{L}_{F}=\frac{\partial \mathcal{L}}{\partial F}.$ Similarly,
variation of the action with respect to $g_{\mu \nu }$ yields the
Einstein-nonlinear Maxwell equations given by%
\begin{equation}
G_{\mu }^{\nu }+\frac{1}{3}\Lambda \delta _{\mu }^{\nu }=8\pi T_{\mu }^{\nu
},
\end{equation}%
where the energy-momentum tensor is 
\begin{equation}
T_{\ \nu }^{\mu }=\frac{1}{2}\left( \delta _{\ \nu }^{\mu }\mathcal{L-}%
4\left( F_{\nu \lambda }F^{\ \mu \lambda }\right) \mathcal{L}_{F}\right) .
\end{equation}%
With these choices we wish now to prove a \textbf{Theorem: } A static
circularly symmetric space time in $2+1-$dimensions having a $U(1)$ electric
field defined in a NED theory necessarily admits a black hole (BH) solution
(The higher dimensional version of this theorem has been considered in Ref. 
\cite{7}).

\textbf{Proof: }For this theorem we choose an electric field two-form as 
\begin{equation}
\mathbf{F}=E\left( r\right) dt\wedge dr.
\end{equation}%
and from (3) one obtains%
\begin{equation}
\sqrt{\left\vert F\right\vert }\mathcal{L}_{F}=\frac{Q}{r},
\end{equation}%
in which $Q$ is an integration constant. The energy-momentum tensor given in
(5) can be expressed alternatively in the form%
\begin{equation}
T_{\ \nu }^{\mu }=\text{diag}\left[ \frac{1}{2}\mathcal{L}-\mathcal{L}_{F}F,%
\frac{1}{2}\mathcal{L}-\mathcal{L}_{F}F,\frac{1}{2}\mathcal{L}\right] .
\end{equation}%
One may also show that the nonzero components of the Einstein tensor are 
\begin{equation}
G_{\mu }^{\nu }=diag\left[ \frac{f^{\prime }}{2r},\frac{f^{\prime }}{2r}.%
\frac{f^{\prime \prime }}{2}\right] 
\end{equation}%
The $rr$ / $tt$ component of the Einstein-Maxwell equation yields the
solution for $f(r)$ in the form%
\begin{equation}
f=-M-\frac{1}{3}\Lambda r^{2}+16\pi \int^{r}r^{\prime }\left( \frac{1}{2}%
\mathcal{L}-\mathcal{L}_{F}F\right) dr^{\prime }
\end{equation}%
in which $M$ is another integration constant. The latter solution satisfies
the $\varphi \varphi $ component of the EM equation. This can be shown by
writing the $\varphi \varphi $ equation as%
\begin{equation}
\frac{f^{\prime \prime }}{2}+\frac{1}{3}\Lambda =4\pi \mathcal{L},
\end{equation}%
but since $f^{\prime \prime }=\left( -\frac{2}{3}\Lambda r+16\pi r\left( 
\frac{1}{2}\mathcal{L}-\mathcal{L}_{F}F\right) \right) ^{\prime }$ it
amounts to%
\begin{equation}
\left( r\left( \mathcal{L}-2\mathcal{L}_{F}F\right) \right) ^{\prime }=%
\mathcal{L}.
\end{equation}%
After some manipulation this is cast into the form%
\begin{equation}
r\left( \mathcal{L}-2\mathcal{L}_{F}F\right) ^{\prime }=2\mathcal{L}%
_{F}F\rightarrow r\left( \mathcal{L}_{F}F^{\prime }-\left( 2\mathcal{L}%
_{F}F\right) ^{\prime }\right) =2\mathcal{L}_{F}F
\end{equation}%
Now, upon dividing by $r\mathcal{L}_{F}F$ and taking the integral it yields $%
\sqrt{\left\vert F\right\vert }\mathcal{L}_{F}=\frac{C}{r}$ which is nothing
but the solution of the Maxwell equation (7) (with $C=Q)$. For appropriate
choices of physical parameters (i.e. $M$, $\Lambda $ and $Q$) the equation $%
f(r)=0$, admits real root(s) and this fact renders the proof of the stated
Theorem valid, irrespective of the choice of $\mathcal{L}\left( F\right) $. 

To get a magnetic dual solution \cite{6} we set 
\begin{equation}
t=i\Phi \text{ \ and \ \ }\varphi =iT
\end{equation}%
which leads from (2) to 
\begin{equation}
ds^{2}=-r^{2}dT^{2}+\frac{1}{f\left( r\right) }dr^{2}+f\left( r\right) d\Phi
^{2},
\end{equation}%
so that the field $2-$form (6) becomes%
\begin{equation}
\mathbf{F}=B\left( r\right) d\Phi \wedge dr
\end{equation}%
in which we have defined%
\begin{equation}
B\left( r\right) =iE\left( r\right) .
\end{equation}%
The metric solution given in (10) is the same metric function $f(r)$ in the
line element (15). It is clear that the solution given here can not
represent a black hole. In addition if we consider $r=r_{0}$ to be the
largest root for $f(r)=0$ in (15), to remove the singularity from the
metric, we introduce 
\begin{equation}
x^{2}=r^{2}-r_{0}^{2}
\end{equation}%
and therefore the line element becomes%
\begin{equation}
ds^{2}=-\left( x^{2}+r_{0}^{2}\right) dT^{2}+\frac{x^{2}}{f\left( x\right)
\left( x^{2}+r_{0}^{2}\right) }dx^{2}+f\left( x\right) d\Phi ^{2}.
\end{equation}%
It is trivial to state that if there is no root for $f(r)=0,$ this
transformation is redundant and therefore $x=r$ with $r_{0}=0$.\ \ \ \ \ \ \
\ \ \ \ \ \ \ \ \ \ \ \ \ \ \ \ \ \ \ \ \ \ \ \ \ \ \ \ \ \ \ \ \ \ \ \ \ \
\ \ \ \ \ \ \ \ \ \ \ \ \ \ \ \ \ \ \ \ \ \ \ \ \ \ \ \ \ \ \ \ \ \ \ \ \ \
\ \ \ \ \ \ \ \ \ \ \ \ \ \ \ \ \ \ \ \ \ \ \ \ \ \ \ \ \ \ \ \ \ \ \ \ \ \
\ \ \ \ \ \ \ \ \ \ \ \ \ \ 

\section{Rotating Charged Black Hole}

To obtain the rotating version of the charged BH solution from a static one
we apply the linear transformation \cite{8} 
\begin{eqnarray}
t &=&\zeta \tilde{t}-\frac{\sigma }{\alpha ^{2}}\tilde{\varphi}, \\
\varphi  &=&\zeta \tilde{\varphi}-\sigma \tilde{t},
\end{eqnarray}%
in which $\zeta ,\sigma $ and $\alpha $ are constant parameters. After the
transformation the line element (2) takes the form%
\begin{equation}
ds^{2}=-\left( N^{0}\right) ^{2}d\tilde{t}^{2}+\frac{dr^{2}}{\tilde{f}^{2}}%
+H^{2}\left( d\tilde{\varphi}+N^{\varphi }d\tilde{t}\right) ^{2}
\end{equation}%
in which%
\begin{eqnarray}
\left( N^{0}\right) ^{2} &=&\left( f\zeta ^{2}-r^{2}\sigma ^{2}+\frac{\left(
r^{2}\zeta \sigma -f\frac{\sigma }{\alpha ^{2}}\zeta \right) ^{2}}{\left(
\zeta ^{2}r^{2}-f\frac{\sigma ^{2}}{\alpha ^{4}}\right) }\right) , \\
H^{2} &=&\left( \zeta ^{2}r^{2}-f\frac{\sigma ^{2}}{\alpha ^{4}}\right) , \\
\tilde{f}^{2} &=&f,
\end{eqnarray}%
and%
\begin{equation}
N^{\varphi }=-\frac{r^{2}\zeta \sigma -f\frac{\sigma }{\alpha ^{2}}\zeta }{%
\zeta ^{2}r^{2}-f\frac{\sigma ^{2}}{\alpha ^{4}}}.
\end{equation}%
We notice that with $\sigma =0$ and $\zeta =1$ (i.e. the identity
transformation) we get the original static line element (2). Furthermore,
upon the transformations (20) and (21), the Maxwell field becomes 
\begin{equation}
\mathbf{F}=E\left( r\right) \left( \zeta d\tilde{t}\wedge dr-\frac{\sigma }{%
\alpha ^{2}}d\tilde{\varphi}\wedge dr\right) ,
\end{equation}%
which clearly shows a magnetic field proportional to the electric field
i.e., $B\left( r\right) =\frac{\zeta \alpha ^{2}}{\sigma }E\left( r\right) .$

By applying the duality principle \cite{6}, given by 
\begin{equation}
\tilde{t}=i\Phi ,\text{ \ \ }\tilde{\varphi}=iT
\end{equation}%
to the metric (22), we obtain 
\begin{equation}
ds^{2}=-H^{2}\left( dT+N^{\varphi }d\Phi \right) ^{2}+\frac{dr^{2}}{\tilde{f}%
}+\left( N^{0}\right) ^{2}d\Phi ^{2}.
\end{equation}%
The limit $\sigma =0$ and $\zeta =1$ yields naturally the line element (15).

\section{An Illustrative Example of NED}

\subsection{Logarithmic Lagrangian In Flat Spacetime}

The action in $2+1-$dimensional flat spacetime with geometrized units \cite%
{6} $G=c=1$ is written as%
\begin{equation}
S=\frac{1}{16\pi }\int dx^{3}\sqrt{-g}\mathcal{L}\left( F\right) ,
\end{equation}%
in which $F=F_{\mu \nu }F^{\mu \nu }$ is the maxwell invariant and \cite{5} 
\begin{equation}
\mathcal{L}\left( F\right) =-\frac{1}{\beta ^{2}}\ln \left( 1+\beta
^{2}F\right) .
\end{equation}%
Herein $\beta $ is a Born-Infeld-like parameter with the dimension of length
such that $\lim_{\beta \rightarrow 0}\mathcal{L}\left( F\right) =-F$ is the
Maxwell limit and $\lim_{\beta \rightarrow \infty }\mathcal{L}\left(
F\right) =0$ is the zero field limit of the Lagrangian. From the outset let
us note that our choice of the source is only electric field with its $2-$%
form given by%
\begin{equation}
\mathbf{F}=E\left( r\right) dt\wedge dr.
\end{equation}%
Variation of the action with respect to the potential $A_{\mu }$ yields the
Maxwell equation (3) in flat space. The flat line element in the circularly
symmetric form is chosen%
\begin{equation}
ds^{2}=-dt^{2}+dr^{2}+r^{2}d\varphi ^{2},
\end{equation}%
and upon substitution of $F=F_{\mu \nu }F^{\mu \nu }=-2E\left( r\right) ^{2}$%
, the Maxwell equation becomes%
\begin{equation}
\frac{\partial \left[ rE\left( r\right) \mathcal{L}_{F}\right] }{\partial r}%
=0\text{.}
\end{equation}%
Simple manipulation yields%
\begin{equation}
\frac{rE}{1-2\beta ^{2}E^{2}}=C
\end{equation}%
in which $C$ is an integration constant. The latter equation admits two
solutions for the electric field, namely 
\begin{equation}
E_{\pm }=-r\frac{1\mp \sqrt{1+\frac{8C^{2}\beta ^{2}}{r^{2}}}}{4C\beta ^{2}}.
\end{equation}%
It should be noted that only the negative branch has the correct Maxwell
limit as%
\begin{equation}
\lim_{\beta \rightarrow 0}E_{-}=\frac{C}{r}
\end{equation}%
which suggests also that we set $C=Q$ as the electric charge. Let us note
that in the sequel we shall exclude the exotic branch $E_{+}$ and choose 
\begin{equation}
E=E_{-}=\frac{r\left( \sqrt{1+\frac{8Q^{2}\beta ^{2}}{r^{2}}}-1\right) }{%
4Q\beta ^{2}}.
\end{equation}%
The Taylor expansion of the electric field in terms of $\beta $ reads%
\begin{equation}
E=\frac{Q}{r}-\frac{2Q^{3}}{r^{3}}\beta ^{2}+8\frac{Q^{5}}{r^{5}}\beta ^{4}+%
\mathcal{O}\left( \beta ^{6}\right) 
\end{equation}%
which obviously gives the correct Maxwell limit. It is observed that the
largest correction to the electric field for small values of $\beta $ is
proportional to $\frac{Q^{3}}{r^{3}}.$ It is also remarkable to observe that
for non-zero $\beta $ the electric field is not singular any more and its
maximum value occurs at $r=0,$ such that $E_{Max}=E\left( r=0\right) =\frac{1%
}{\sqrt{2}\left\vert \beta \right\vert }.$ The exact electric potential is
also given by 
\begin{equation}
V-V_{ref}=-\int Edr=-Q\ln r-Q\ln \left( 1+\sqrt{1+\frac{8Q^{2}\beta ^{2}}{%
r^{2}}}\right) -\frac{r^{2}}{8Q\beta ^{2}}\left( \sqrt{1+\frac{8Q^{2}\beta
^{2}}{r^{2}}}-1\right) 
\end{equation}%
in which $V_{ref}$ is an integration constant fixed by $V_{ref}=Q\left( 
\frac{1}{2}+\ln 2\right) $ to yield the Maxwell limit. Expansion of the
electric potential reads as%
\begin{equation}
V=-Q\ln r-\frac{Q^{3}}{r^{2}}\beta ^{2}+2\frac{Q^{5}}{r^{4}}\beta ^{4}+%
\mathcal{O}\left( \beta ^{6}\right) 
\end{equation}%
which is in conform with (39). Here also one can show that the electric
potential is not singular any more and its value at $r=0$ is given by%
\begin{equation}
V\left( r=0\right) =-\frac{Q}{2}\left( 1-\ln \left( 2Q^{2}\beta ^{2}\right)
\right) .
\end{equation}

\subsection{Logarithmic Lagrangian in Curved Spacetime}

In the curved spacetime with the extended Maxwell Lagrangian (31) coupled
with the gravity via the action (1) and an electric field ansatz given in
(6), the Einstein-Maxwell equations admit the electric field found in (36)
and the metric solution 
\begin{equation}
f\left( r\right) =-\tilde{M}+\frac{r^{2}}{\tilde{\ell}^{2}}-\frac{\left[
3r^{2}\sqrt{1+\frac{8Q^{2}\beta ^{2}}{r^{2}}}+8Q^{2}\beta ^{2}\ln \left( r%
\left[ \sqrt{1+\frac{8Q^{2}\beta ^{2}}{r^{2}}}+1\right] \right) +2r^{2}\ln
\left( r^{2}\left[ \sqrt{1+\frac{8Q^{2}\beta ^{2}}{r^{2}}}-1\right] \right) %
\right] }{4\beta ^{2}}.
\end{equation}%
Herein 
\begin{equation}
\tilde{M}=M+Q^{2}\left[ \frac{\ln \left( 2Q^{2}\right) -2}{\left( 1+\beta
^{2}\right) }-\ln \left[ 8Q^{2}\left( 1+\beta ^{2}\right) \right] \right] 
\end{equation}%
and 
\begin{equation}
\frac{1}{\tilde{\ell}^{2}}=\frac{1}{\ell ^{2}}+\frac{\frac{3}{4}+\ln 2+\ln
\left\vert Q\beta \right\vert }{\beta ^{2}}.
\end{equation}%
Further, $\frac{1}{\ell ^{2}}=-\frac{\Lambda }{3}$ and $M$ is the usual mass
of the BTZ black hole i.e., 
\begin{equation}
\lim_{\beta \rightarrow 0}f\left( r\right) =-M+\frac{r^{2}}{\ell ^{2}}%
-Q^{2}\ln r^{2}
\end{equation}%
and 
\begin{equation}
\lim_{\beta \rightarrow \infty }f\left( r\right) =-M+\frac{r^{2}}{\ell ^{2}}
\end{equation}%
as expected.

\subsection{Magnetic Solution}

In accordance with the duality transformation (14) and $Q=iP$ a magnetic
solution and the corresponding Maxwell's field $2-$form are given by (15)
and (16) respectively in which 
\begin{equation}
B\left( r\right) =\frac{r\left( \sqrt{1-\frac{8P^{2}\beta ^{2}}{r^{2}}}%
-1\right) }{4P\beta ^{2}},
\end{equation}%
and the metric function reads as%
\begin{equation}
f\left( r\right) =-\tilde{M}+\frac{r^{2}}{\tilde{\ell}^{2}}-\frac{\left[
3r^{2}\sqrt{1-\frac{8P^{2}\beta ^{2}}{r^{2}}}-8P^{2}\beta ^{2}\ln \left( r%
\left[ \sqrt{1-\frac{8P^{2}\beta ^{2}}{r^{2}}}+1\right] \right) +2r^{2}\ln
\left( r^{2}\left[ \sqrt{1-\frac{8P^{2}\beta ^{2}}{r^{2}}}-1\right] \right) %
\right] }{4\beta ^{2}}.
\end{equation}%
Once more we note that in order to resolve the issue of any possible
coordinate singularity one may use the transformation given in (18) to get
the free coordinate singularity metric as given in (19) which in contrast to
the electric case this is clearly not a black hole solution.

\section{Conclusion}

Since its inception with the Born-Infeld version the basic advantage of NED
is to eliminate the singularities. Herein we apply the same rule to the $2+1-
$dimensional BTZ black hole in the presence of a logarithmic Lagrangian as
representative of the NED. The results are interesting: Unlike the linear
Maxwell theory the electric field, and the electric potential are
well-defined and regular whereas the spacetime persists to be singular at $%
r=0$. In a separate study we describe a new method to construct regular
black holes in $2+1-$dimensions \cite{9}. This makes use of the additional
Hoffmann-Infeld term in the NED to eliminate the singularity at $r=0$ in a
natural way based on cutting and pasting technique. A Theorem is proved in
connection with $2+1-$dimensions, that the spacetime must admit black holes
for certain choices of physical parameters (i.e., $M$, $\Lambda $ and $Q$)
whenever electric fields are involved. No doubt, this result covers also the
pure electrically charged linear Maxwell fields. It is also implied
indirectly, by appealing to a duality principle of $2+1-$dimensional
spacetime that the magnetic fields do not create black holes. This seems to
be one of the peculiar features of electromagnetism, whether linear or
nonlinear, in $2+1-$dimensional spacetime.

\end{document}